\documentclass[%
aps,
 amsmath,amssymb,
preprint,%
]{revtex4-1}

\usepackage{graphicx}
\usepackage{dcolumn}
\usepackage{bm}
\usepackage{hyperref}
\hypersetup{
    colorlinks=true,
    linkcolor=blue,
    filecolor=magenta,      
    urlcolor=blue,
    citecolor=blue,
}
\usepackage{color}

\begin{document}



\title{Enhanced Electron Transport in Thin Copper Films via Atomic-Layer Materials Capping}%

\author{David M. Guzman}
\author{Alejandro Strachan}
 \email[Corresponding Author: ]{strachan@purdue.edu}
\affiliation{School of Materials Engineering and Birck Nanotechnology Center, Purdue University, West Lafayette, Indiana 47907, USA}

\date{\today}

\begin{abstract}
Using first-principles calculations based on density functional theory and non-equilibrium Green's functions, we characterized the effect of surface termination on the electronic transport properties of nanoscale Cu slabs. With ideal, clean (111) surfaces and oxidized ones as baselines we explore the effect of capping the slabs with graphene, hexagonal boron nitrate, molybdenum disulfide and stanene. Surface oxide  suppresses balistic conductance by a factor of 10 compared to the ideal surface. Capping the ideal copper surface with graphene slightly increase conductance but MoS$_2$ and stanene have the opposite effect due to stronger interactions at the interface. Interestingly, we find that capping atomistically roughed copper surfaces with graphene or MoS$_2$ decreases the resistance per unit length by 20 and 13\%, respectively, due to reduced scattering. The results presented in this work suggest that two-dimensional materials can be used as an ultra-thin liner in metallic interconnect technology without increasing the interconnect line resistivity significantly.
%
\end{abstract}


\maketitle

\section{Introduction}
\label{sec:intro}

The downscaling of integrated circuits is limited, among other factors, by increasing power dissipation in metallic interconnects. As the node size continues to shrink, the linewidth of copper interconnects has to scale accordingly, causing a substantial increase in resistivity\cite{Havemann2001,Kapur2002,Graham2010}. Different scattering mechanisms, such as electrons scattered by phonons, impurity ions, grain boundaries, and surfaces,  contribute to the total resistivity. As the cross-section of the conductor becomes comparable to the electron mean free path (approximately 40 nm at room temperature), surface scattering becomes increasingly important; furthermore, downscaling also results in grain refinement and, consequently, increased grain boundary scattering. These size effects become evident when the width of the interconnect falls below 150 nm, roughly three times the electron mean free path, and become critical bellow 75 nm where the resistivity of the interconnect is approximately 3.2 $\mu\Omega\cdot cm$ at 300K, almost twice that of bulk copper at same temperature\cite{Plombon2006}. Even though debate still exists regarding the impact of individual scattering mechanism on the overall resistivity of metallic interconnects, several theoretical\cite{Ke2009,Valencia2017,Cesar2014,Chawla2011,Timoshevskii2008} and experimental\cite{Barnat2002,Liu2001,Purswani2007} studies suggest that understanding the role of surface scattering is critical to design the next generation of interconnect technology.

Surface electronic states are generated by the abrupt transition from bulk solid material to a surface termination and are localized around atoms within a few atomic layers from the surface. In metals with clean and ideal surface terminations, surface states, also known as Shockley states\cite{Shockley1939,Schiller2006}, emerge from the decoupling of a surface state from the bulk band states\cite{Forstmann1993} and exhibit nearly free electron characteristics confined in a two-dimensional space. Shockley states significantly affect surface properties such as catalytic reactions\cite{Straube2000} and adsorption mechanisms\cite{Yan2015}. Thus, surface chemistry and fabrication methods largely determine electron scattering at interconnect surfaces and can modify its character between entirely specular (elastic) and diffusive (inelastic). Atomistically flat free copper surfaces exhibit specular scattering and degrade properties very significantly\cite{Timoshevskii2008,Ke2009}; however, in practical applications, metallic surfaces are exposed to environments that modify the surface features far from their ideal configuration resulting in increased surface scattering, hence reduced thermal and electrical conductivity\cite{Rossnagel2004}. For instance, exposure of copper surfaces to oxidizing environments cause the formation of non-conducting thin copper oxide layers\cite{Platzman2008} which modifies the character of the surface scattering, thus increasing the overall resistivity of the interconnect. Successful fabrication of oxidation resistant copper films with high conductivity via surface alloying has been previously reported\cite{Ding1994,Li2016,Peng2017}; however, these innovations have resulted in materials that are not scalable and exhibit high electrical resistance for the interconnect technology standard and as a consequence not suitable for this application. 

Besides the increased electrical resistivity of metallic thin films, copper interconnect technology faces reliability issues concerning the tendency of copper to diffuse into the dielectric material that separates interconnect lines causing undesired electrical connections; furthermore, diffusion of copper into the transistors could cause integrated circuit failure\cite{Shacham-Diamand1993}. As a consequence, the copper interconnect must be surrounded by a diffusion barrier (liner) to avoid the direct interaction of copper with the dielectric material preventing metal diffusion. Ta, TaN, and TiN have been used as traditional liner materials\cite{Li2004} while others, such as Ru/TaN(Ti)\cite{Yang2010} and CuMn alloys\cite{Watanabe2007}, have been proposed as effective diffusion barriers. However, these liner materials exhibit resistivity that is significantly higher than that of copper\cite{Kapur2002}; therefore, to maintain the interconnect resistivity to the standards, the liner needs to be ultra-thin compromising their diffusion barrier effectiveness. On the other hand, while liner materials are selected based on, among other criteria, low chemical reactivity with copper, the interactions are strong enough to introduce surface defects that enhance diffusive electron surface scattering increasing the interconnect resistivity\cite{Hinode2001,Kitada2009}.  

Two-dimensional materials, such as graphene, hexagonal boron nitride, and transition metal dichalcogenides have been recently suggested as a promising diffusion barrier and oxidation resistance alternatives\cite{Chen2011,Liu2013,Kang2013,Mehta2015,Lo2017} due to their exceptional structural, mechanical and electronic properties. Mehta \textit{et al.}\cite{Mehta2015} found that graphene encapsulated copper nanowires exhibit 7 to 15\% higher electrical conductivity than air-exposed Cu nanowires (oxidized surfaces) and enhanced thermal transport properties\cite{Goli2014}. Another experimental study\cite{Lo2017} demonstrated the use h-BN and MoS$_2$ in standard damascene structures as effective ultra-thin copper diffusion barriers, and a recent \textit{ab initio} study\cite{Cuong2017} showed that graphene and h-BN provides an oxidation protectant coating to pristine copper surfaces and that Shockley surface states with 2D electron gas-like features persist after capping the clean copper surfaces. However, the electronic transport properties of defective copper surfaces encapsulated with atomic-layer materials (Cu-hybrids) remains unexplored. Specifically, we lack an understanding of how the nature of the 2D material and its interaction with Cu (including possible defects) affect transport.

In this paper, we present an \textit{ab initio} study based on density functional theory (DFT) of the electronic transport properties of Cu-hybrids capped with graphene (Gr), hexagonal boron nitride (h-BN), molybdenum disulfide (MoS$_2$), stanene, and oxidized copper surface. We find that the weak interaction of graphene and h-BN with copper leave the surface electronic structure unchanged; hence the conductance of the bare pristine copper slabs and that of the Cu-hybrid is similar. In the case of Cu/Gr hybrids, the shift in Fermi energy accounts for extra conduction channel in the graphene sheets making the hybrid more conductive compared to the bare Cu slab. On the other hand, MoS$_2$, stanene, and oxidized surfaces exhibit stronger interactions with the copper surfaces allowing for stronger surface scattering which is detrimental to the hybrid conductance. Lastly, the length dependent resistance of defective copper surfaces and Cu-hybrids exhibit, as expected, a linear trend and we find that the resistance per unit length of Cu/Gr and Cu/MoS$_2$ is reduced by 20\% and 12\%, respectively.

The remainder of this paper is organized as follows: Section \ref{sec:methods} discusses the models, simulation details, and systems energetics. In Section \ref{sec:ElecTransI} we present and discuss the electron transport on pristine Cu-hybrids, followed by the length dependent transport properties in Section \ref{sec:ElecTransII}. Our conclusions are presented in Section \ref{sec:Conc}.

\newpage

\section{Methods}
\label{sec:methods}

\subsection{Atomistic Models} 

We characterized the electrical conductivities of a series of systems: bulk Cu, thin slabs with perfect surfaces and slabs with surfaces containing vacancies, in addition, we studied slabs with oxidized surfaces and slabs capped with 2D materials (graphene, h-BN, MoS$_2$, and stanene. All slabs are oriented with low energy (111) free surfaces and a thickness of 7 atomic layers, approximately 1.5 nm. The atomic-layer materials are added to the top and bottom copper surfaces in a supercell arrangement that minimizes the absolute strain on the system by choosing the copper and capping material in-plane cell vectors $\vec{T}_{Cu}$ and  $\vec{T}_{Cap}$ in such a way that it minimizes $\delta=\frac{||\vec{T}_{Cap}-\vec{T}_{Cu}||}{||\vec{T}_{Cu}||}$ with the constraint that only the capping material is allowed to rotate, and the resultant supercell is orthorhombic. The method is described in Refs \cite{Farmanbar2016,Komsa2013,Helfrecht2017} and all model details are listed in Table \ref{tab:models}.

Oxidized copper surfaces are generated using the melt-and-quench method using molecular dynamics with the reactive force field ReaxFF\cite{VanDuin2010a}. An initial crystal layer of CuO is melted using MD at T=1200 K while maintaining the Cu slab at T=300K. The molten oxide is then cooled down to 300K at a rate of 50K/ps. The resulting structures exhibit an amorphous oxide layer, about 0.5 nm in thickness and resemble experimental interfaces\cite{Matsumoto2001,Soon2006}. 

\begin{table}[!ht]
    \centering
    \caption{Copper hybrid simulation models details}
    \label{tab:models}
    \begin{tabular}{|l|c|c|c|c|}
        \hline
        Hybrid Structure  & Number of Atoms & Absolute Strain [\%] & $|\vec{A}|$ [\AA] & $|\vec{B}|$ [\AA] \\ \hline
        Cu/Graphene      & 44              & 1.2                   & 2.51             & 8.69              \\ \hline
        Cu/h-BN              & 44              & 0.6                   & 2.52             & 8.72              \\ \hline
        Cu/MoS$_2$       & 132             & 2.0                   & 5.29             & 12.95             \\ \hline
        Cu/Stanene         & 50              & 0.8                   & 4.37             & 7.77              \\ \hline
        Cu/CuOx             & 216             & ---                   & 8.81             & 15.45             \\ \hline
    \end{tabular}
\end{table}

\begin{figure}[!ht]
\centering
\includegraphics[width=0.65\textwidth]{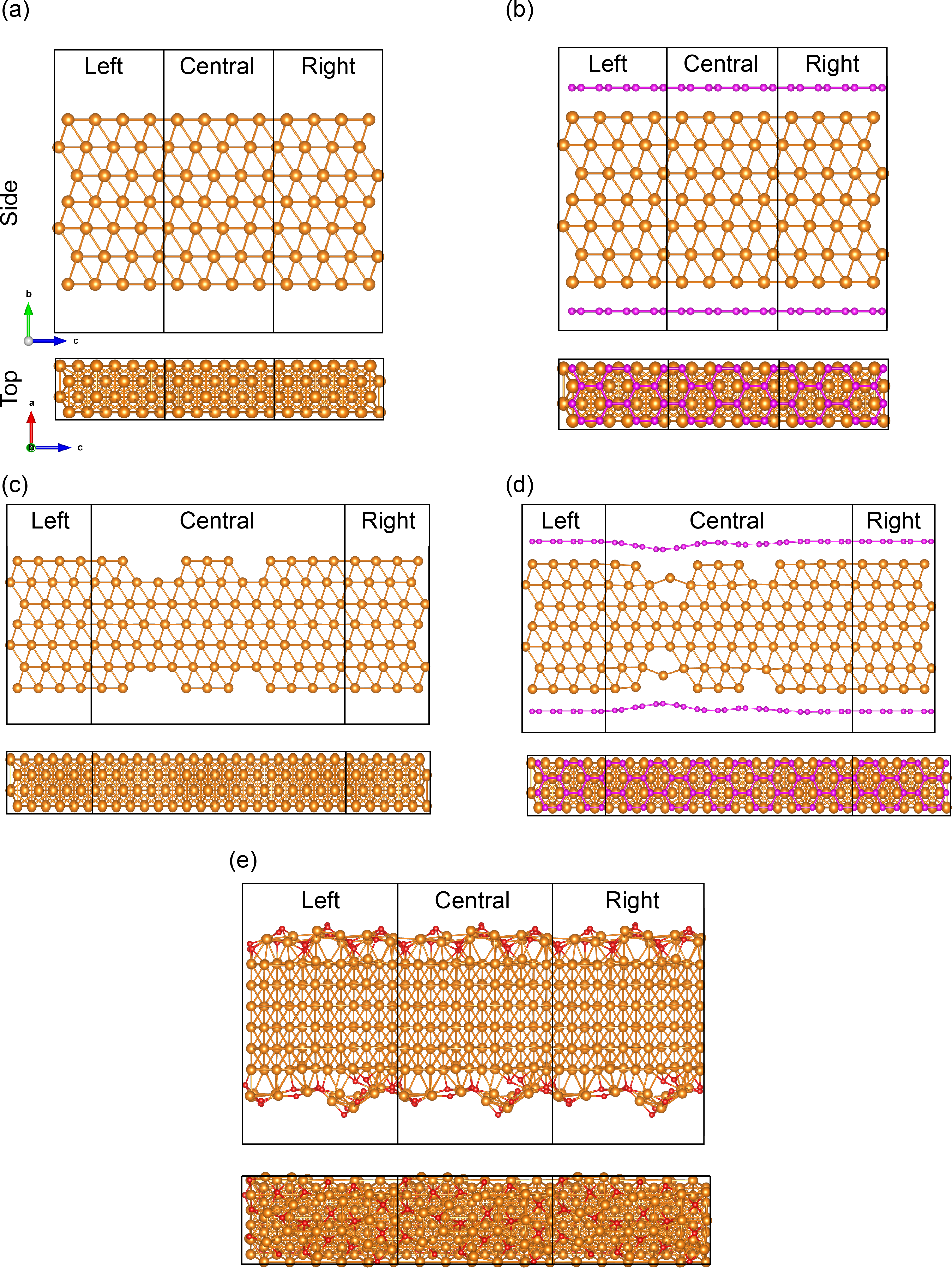}
\caption{Side and top views of the optimized two-probe devices for (a) copper slab, (b) Cu/Graphene hybrid, (c) Cu slab with surface defects, (d) Cu/Graphene hybrid with surface defects, and (e) oxidized copper surface.}
\label{fig:struc}
\end{figure}

\subsection{Density Functional Theory}
The structural optimization and energetic calculations were carried out using density functional theory as implemented in the Vienna \textit{ab initio} simulation package (VASP)\cite{Kresse:1996kg,Kresse:1996kl} using projector-augmented wave (PAW) pseudopotentials\cite{Blochl:1994dx,Kresse:1999dk}. The generalized gradient approximation (GGA) as proposed by Perdew-Burke-Ernzerhof (PBE) \cite{Perdew:1996iq} was used as the exchange-correlation functional. Since long-range van der Waals interactions are not well captured by Born- Oppenheimer–based DFT, we employed Grimme's DFT-D3 energy correction scheme to account for these dispersion interactions\cite{Grimme:2006fca,Grimme:2010ij}. The plane wave-functions are expanded to a kinetic energy cutoff of 500 eV. For structural relaxations, we used $\Gamma$-centered \textit{k} grids of sizes 4 $\times$ 4 $\times$ 1 to 8 $\times$ 8 $\times$ 1, depending on the size of the supercell, resulting in linear grid spacings of 0.10 - 0.15 \AA$^{-1}$. The self-consistent field routine was terminated when the energy change between cycles fell below 10$^{−5}$ eV, and structural relaxations were terminated when all forces fell below 0.01 eV/\AA.

\subsection{Electron Transport - Non-equilibrium Green's Functions} 
The electronic transport is treated in the ballistic regime within the phase-coherent approximation of self-consistent non-equilibrium Green's function (NEGF) method. All electronic transport computations are carried out using the TranSiesta code\cite{Brandbyge:2002ck}, implemented in the SIESTA package\cite{Ordejon:1996je,Soler:2002kr}. The core electrons are replaced by pseudopotentials, and the valence electrons are represented by basis sets made up of numerical atomic-like wave-functions. In this work we use norm-conserving Troullier-Martins pseudopotentials\cite{Troullier:1990dy} for all atomic species and the exchange-correlation potential is calculated within the GGA-PBE\cite{Perdew:1996iq} functional. Double zeta plus polarization (DZP) numerical orbital basis sets are used for all atomic species.

The device model for the electronic transport adopts a two-probe configuration as shown in Figure \ref{fig:struc}. The self-consistent field calculation was terminated when an energy tolerance of 10$^{−5}$ eV is reached. \textit{k}-mesh of 4 $\times$ 1 $\times$ 100 was used for the computation of the electrodes self-energy while the transmission spectra calculations was carried out on a 80 $\times$ 1 $\times$ 1 \textit{k}-grid to ensure converged values.

\textcolor{black}{Our simulations of the two-probe devices consist of a left-lead(L), a central-region(C), and a right-lead(R), where the self-consistent Kohn-Sham potentials and Hamiltonian matrices of the leads come from a separate fully periodic DFT calculation. The retarded Green's function is only calculated in the central region, which is treated as an open system, using the Hamiltonian of the central region (H$_C$) and the electrode self-energies ($\Sigma_{L}$ and $\Sigma_{R}$) according to Equation \ref{eq:green},}
\begin{equation}
	G^{ret}=\left[(\varepsilon+i\eta_+)S - H - \Sigma_{L}(\varepsilon)-\Sigma_{R}(\varepsilon)\right]^{-1}
	\label{eq:green}
\end{equation}
\textcolor{black}{where S is the the overlap matrix and $\eta_+$ is an infinitesimal positive number. The left and right density matrix contributions are calculated as:}
\begin{equation}
	D_{L,R}=\int\rho_{L,R}(\varepsilon)n_F(\varepsilon-\mu_{L,R})d\varepsilon
	\label{eq:densmatrix}
\end{equation}
\textcolor{black}{where the $n_F$ is the Fermi function, $\mu_{L,R}$ the electrochemical potential of the left or right electrode, and the spectral function $\rho(\varepsilon)$ is given by}
\begin{equation}
	\rho_{L,R}(\varepsilon)=\frac{1}{2\pi}G^{ret}(\varepsilon)\Gamma_{L,R}(\varepsilon)G^{adv}(\varepsilon).
	\label{eq:spec}
\end{equation}
\textcolor{black}{In Equation \ref{eq:spec}, $G^{adv}(\varepsilon)$ is the advanced Greens function defined as $G^{adv}(\varepsilon)=(G^{ret}(\varepsilon))^{\dagger}$. And $\Gamma_{L,R}(\varepsilon)$ is known as the broadening function of the left and right electrode given by:}
\begin{equation}
	\Gamma_{L,R}(\varepsilon)=\frac{1}{i}(\Sigma_{L,R}-(\Sigma_{L,R})^{\dagger}).
	\label{eq:broad}
\end{equation}
\textcolor{black}{The electron density of the central region is calculated from the total density matrix $D=D_L+D_R$ as:}
\begin{equation}
	\mathcal{N}(\mathbf{r})=\sum_{ij}D_{ij}\phi_i(\mathbf{r})\phi_j(\mathbf{r}).
	\label{eq:elecdens}
\end{equation}
\textcolor{black}{where $\phi$ are numerical basis orbitals. And the total transmission and total current for an infinitesimal bias are computed as:}
\begin{equation}
	T(\varepsilon)=G^{ret}(\varepsilon)\Gamma_L(\varepsilon)G^{adv}(\varepsilon)\Gamma_R(\varepsilon),
\end{equation}
\textcolor{black}{and}
\begin{equation}
	I=\frac{e}{h}\int_{-\infty}^{\infty} T(\varepsilon)[n_F(\varepsilon-\mu_{L})-n_F(\varepsilon-\mu_{R})]d\varepsilon.
\end{equation}

\textcolor{black}{Following the framework developed by Solomon \textit{et al.}\cite{Solomon2010}, the local currents are defined based on the projection of transmission onto all pair of atoms in the central region. For an infinitesimal bias, the local current from an atom at site $n$ to an atom at site $m$ is written as:}
 \begin{equation}
	I_{mn}^{local}(\varepsilon) = \frac{2e}{h}\sum_{i\in m}\sum_{\substack{j\in n\\ n \notin m}}\int_{-\infty}^{\infty}\frac{1}{2\pi}[V_{ij}G^<_{ji}(\varepsilon) - V_{ji}G^<_{ij}(\varepsilon)]d\varepsilon,
	\label{eq:localI}
\end{equation}
\textcolor{black}{where $V_{ij}$ is the coupling between orbitals $i$ and $j$, and the lesser Green's function $G^(\varepsilon)$ is defined as:}
 \begin{equation}
 G^<(\varepsilon)=G^{ret}(\varepsilon)\Sigma^{<}(\varepsilon)G^{adv}(\varepsilon),
\end{equation}
\textcolor{black}{and the lesser self-energy is written as $\Sigma^<(\varepsilon)=in_F(\varepsilon-\mu_{L})\Gamma^L(\varepsilon)+in_F(\varepsilon-\mu_{R})\Gamma^R(\varepsilon)$. From Equation \ref{eq:localI}, the integrand is recognized as the local transmission given by:}
 \begin{equation}
	T_{mn}^{local}(\varepsilon) = \sum_{i\in m}\sum_{\substack{j\in n\\ n \notin m}}[V_{ij}G^<_{ji}(\varepsilon) - V_{ji}G^<_{ij}(\varepsilon)].
	\label{eq:localT}
\end{equation}

\textcolor{black}{Equation \ref{eq:localT} is the projected transmission onto all pairs of atoms in the central region and provides an useful atomistic representation of the electron transmission modes and their probability. This scheme has been widely used to describe how current flows in molecular electronics, conductive elements, and metal-semiconductor/insulator-metal structures\cite{Deng2014,Srivastava2015,Shen2012,Solomon2010}.}

\subsection{Adhesion Energy}
\label{sec:AE}

After relaxing the structures, we characterized the adhesion energy of all the atomic-layered materials to the copper (111) surface, see table \ref{tab:adhesion}. 
Even though the adhesion energies of these materials are smaller than those involving chemical bonds, they are larger than the binding energies between
layers of the corresponding layered materials \cite{Bjorkman2012} indicating that all hybrid interconnects studied would be stable. These results are also important for
validation purposes and to establish a guideline for the accuracy of our calculations. For example, our DFT+D3 calculated adhesion energy and separation distance of the Cu/Graphene system is 0.5 J/m$^2$ and  3.25 \AA, respectively, underestimates the binding energy of 0.72 J/m$^2$ obtained from delamination experiments on Cu/Graphene systems\cite{Yoon2012}. In the case of h-BN, our calculation of adhesion energy (0.54 J/m$^2$) overestimates the experimental value of 
0.18 J/m$^2$ estimated from surface states dispersion curves\cite{Joshi2012}. In general, our DFT+D3 simulations tend to predict higher adhesion energies and shorter separations distances when compared to other van der Waals corrected exchange-correlation functionals\cite{Becke1988, Dion2004}. For the case of 
Cu/MoS$_2$ other \textit{ab initio} simulations\cite{Farmanbar2016} based on a modified Becke88 functional\cite{Klimes2011} estimates the binding energy and 
separation distance to be 0.40 eV/MoS$_2$ and 2.5 \AA, respectively. This is to be compared to our DFT+D3 results for the binding energy (0.85 eV/MoS$_2$) 
and separation (2.25 \AA). Similarly, DFT calculations based on the modified Becke88 functional predicts the binding energy and separation distance of the 
Cu/Graphene system to be 0.2 J/m$^2$ and 3.26 \AA\cite{Giovannetti2008} and for the Cu/hBN system 0.42 J/m$^2$ and 3.27 \AA\cite{Bokdam2014}.

\begin{table}[!ht]
    \centering
    \caption{Separation distance and adhesion energy of the copper hybrid systems}
    \label{tab:adhesion}
    \begin{tabular}{|l|c|c|c|}
        \hline
        Hybrid Structure & Separation Distance [\AA] & Adhesion Energy [meV/\AA$^2$] & Adhesion Energy [J/m$^2$] \\ \hline
        Cu/Graphene  & 3.25      & 32 & 0.50                                 \\ \hline
        Cu/h-BN      & 3.23    & 30  & 0.54                                  \\ \hline
        Cu/Stanene   & ZZ  & XX     & XX                                \\ \hline
        Cu/MoS2      & 2.25    & 90  & 1.50                                \\ \hline
    \end{tabular}
\end{table}

\newpage

\section{Ballistic Electron Transport on Perfect Surfaces} 
\label{sec:ElecTransI}

\begin{figure}[ht]
    \centering
    \includegraphics[width=1\textwidth]{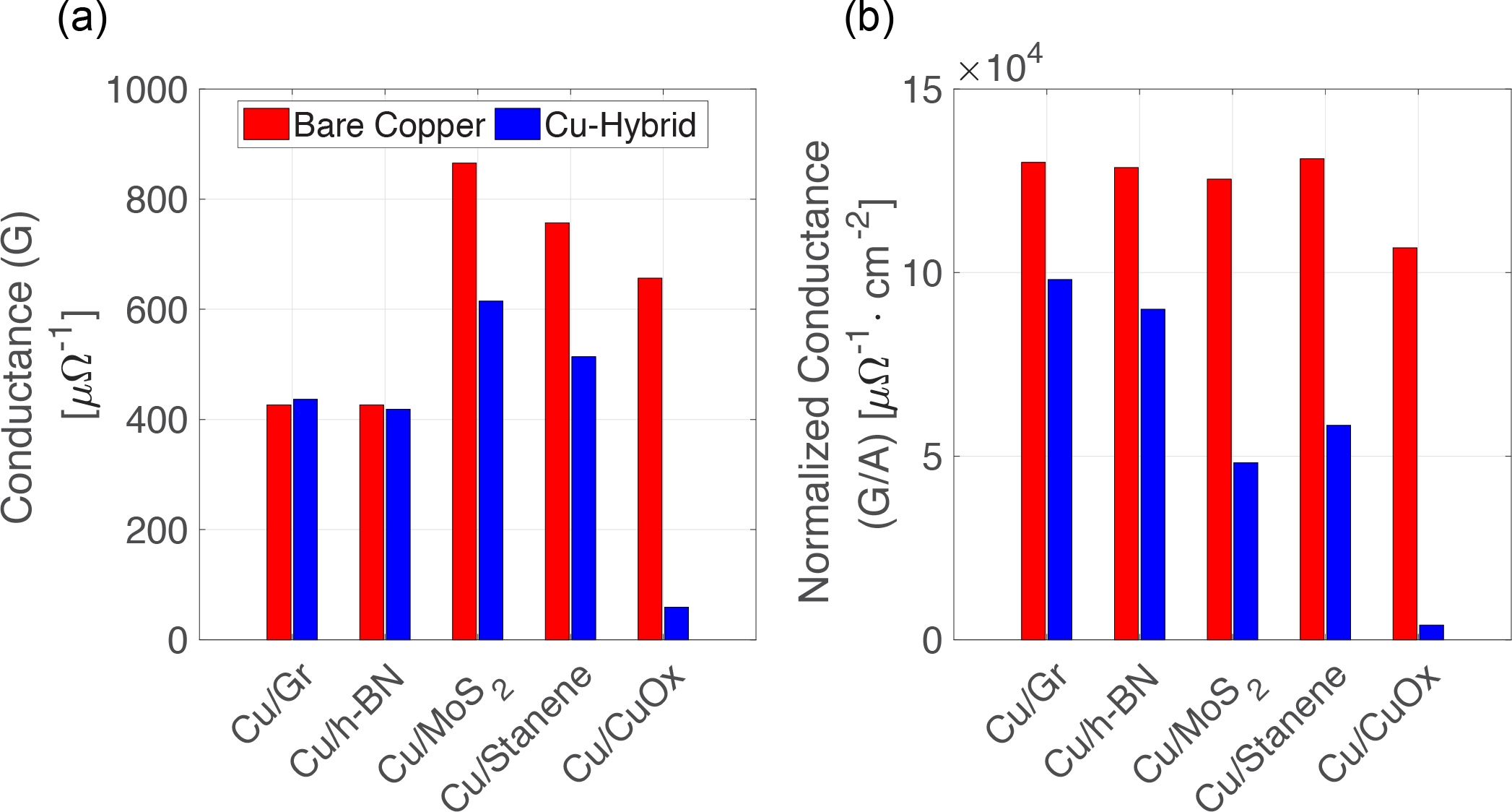}
    \caption{(a) Ballistic conductance of pristine (red bars) and Cu-hybrids (blue bars). (b) Conductance normalized by the cross sectional area.}
    \label{fig:ballistic}
\end{figure}

To characterize the effect of the various surface terminations on electronic transport, we use pristine, defect-free, Cu slabs as a reference. 
Figure \ref{fig:ballistic}a shows the ballistic conductance of the copper hybrids (red) and the corresponding clean Cu slabs (blue). We note
that the transport orientation of all copper thin films is in the [$\bar{1}\bar{1}2$] direction while their width is oriented in the [$\bar{1}10$] direction and the difference in conductance between the reference structures is attributed to the difference in cross-sectional areas. 
In all cases except for the Cu/Graphene hybrid, the presence of surface passivation or capping reduces conductance. Interestingly, adding graphene 
(with no electronic states at the Fermi energy and consequently zero conductance in its free-standing form) increases the conductance of the system.
In all remaining cases, the capping or surface termination reduces conductance; this reduction can be due to a reduction in the number of states available for transport or an increase in surface/interface scattering, reducing the specularity of the Cu surface. 
Not surprisingly, the oxide layer results in a dramatic reduction in conductance and h-BN results in a minor change in conductance due to weak
interactions. Capping Cu with MoS$_2$, which is known to result in a complex re-distribution of the electronic density at the interface\cite{Helfrecht2017}, results in a reduction of the conductance of approximately 30\% and stanene leads to a similar result.

Figure \ref{fig:ballistic}b shows conductance normalized by the area. As expected, the conductance per cross-sectional area of the bare 
copper thin films is approximately the same ($\sim$13$\times$10$^4$ $\mu\Omega/cm^2$), while G/A of the hybrid structures is substantially 
lower than that of the clean copper slabs, which implies that the density of conducting modes in the capping material is very low.

A few interesting cases stand out from Figure \ref{fig:ballistic}a, Cu/Graphene where the conductance of the hybrid is slightly higher than that of the bare copper, 
Cu/h-BN and Cu/MoS$_2$ where despite the semiconducting character of the capping material, the effect on the conductance is very different. 
The conductance of the Cu/h-BN hybrid and perfect Cu slab is the same while that of the Cu/MoS$_2$ hybrid is reduced. 
And the oxidized copper surfaces (Cu/CuOx) where just a thin layer of oxide has the effect of suppressing conductance by a factor of 10.  

\begin{figure}[ht]
    \centering
    \includegraphics[width=0.7\textwidth]{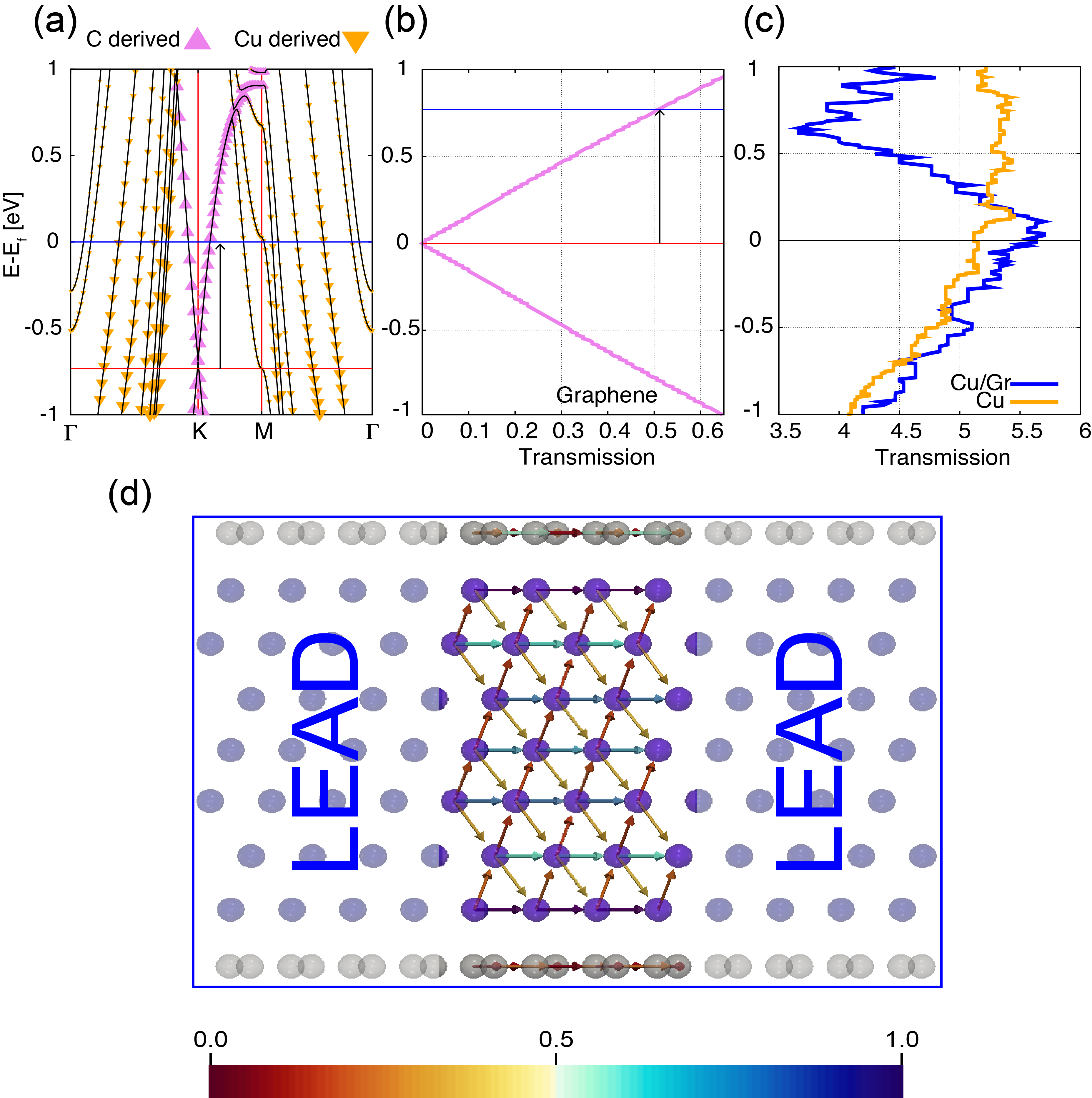}
    \caption{(a) Electronic band structure of the Cu/Graphene hybrid showing the character bands for carbon (violet) and copper (orange). The red line at the Dirac point of graphene and blue line at the Fermi level of the hybrid indicates the change in Fermi energy. (b) Transmission function of graphene. The blue line indicates the position of the Fermi level in the hybrid system. (c) Transmission function of  the Cu/Graphene hybrid (blue) and copper only (orange). (d) Local transmission at the Fermi energy of the Cu/Graphene hybrid. The color bar represents the magnitude of the transmission probability.}
    \label{fig:Cu-Gr}
\end{figure}

{\bf Copper/graphene.} The enhanced conductance in the Cu/Gr hybrid originates from two effects. First, the relatively weak, non-covalent, interaction between copper and graphene does not disturb the electronic structure of copper in a significant manner, hence the absence of additional surface scattering. 
Second, the Fermi energy of the Cu/Graphene hybrid equilibrates approximately 0.75 eV above the Dirac point of graphene, as shown in the electron band 
dispersion of the hybrid structure Figure \ref{fig:Cu-Gr}a where the bands marked with violet up triangles correspond to carbon derived states while those marked with orange down triangles indicate copper derived states. As a consequence, graphene contributes to electron conduction since a few transmission channels become available. 
Figure \ref{fig:Cu-Gr}b shows the \textit{k} averaged transmission coefficient of two free-standing non-interacting graphene as a function of energy. The blue line indicates the 
Fermi energy of the hybrid, the increase in Fermi energy results in a transmission of 0.5 contributed by the non-interacting top and bottom graphene layers (see Figure \ref{fig:struc}b), this can be compared to the increase in transmission when graphene is added to a Cu slab. From the \textit{k} averaged transmission of copper (orange) and the Cu/graphene hybrid (blue) shown in 
Figure \ref{fig:Cu-Gr}c a difference in transmission of about 0.5 can be identified at the Fermi level (zero line), such difference is explained by the conducting channels contribution of graphene.

\textcolor{black}{To further understand the electron transmission modes in the Cu/Graphene hybrids, we computed the local transmission pathways in the central region where the possible transmission pathway between atoms $i$ and $j$ at the Fermi energy is represented by an arrow connecting such atoms. In these plots, a transmission pathway between a pair of atoms is shown if its magnitude if at least 10\% of the largest local transmission. Furthermore, the local transmission pathways are normalized with respect to the largest magnitude of the local-bond contributions, irrespective of the magnitude of the total transmission at that energy; that is, the magnitude of the individual arrows is color coded between 0 (red) and 1 (blue).}

\textcolor{black}{The local transmission pathways in the Cu/Graphene hybrid are shown in Figure \ref{fig:Cu-Gr}d. Consistent with the character bands plot (Fig. \ref{fig:Cu-Gr}a), the transmission pathways are localized within the copper slab and graphene layers only; that is, there is no contributions to the total current from local transmission between the copper surface and graphene layers. This is attributed to the lack of hybridization between carbon and copper atoms due to the weak interaction between these two materials. The absence of local transmission pathways from the copper surfaces to the graphene layers indicates that graphene capping does not introduce additional surface scattering}
 
\begin{figure}[ht]
    \centering
    \includegraphics[width=0.7\textwidth]{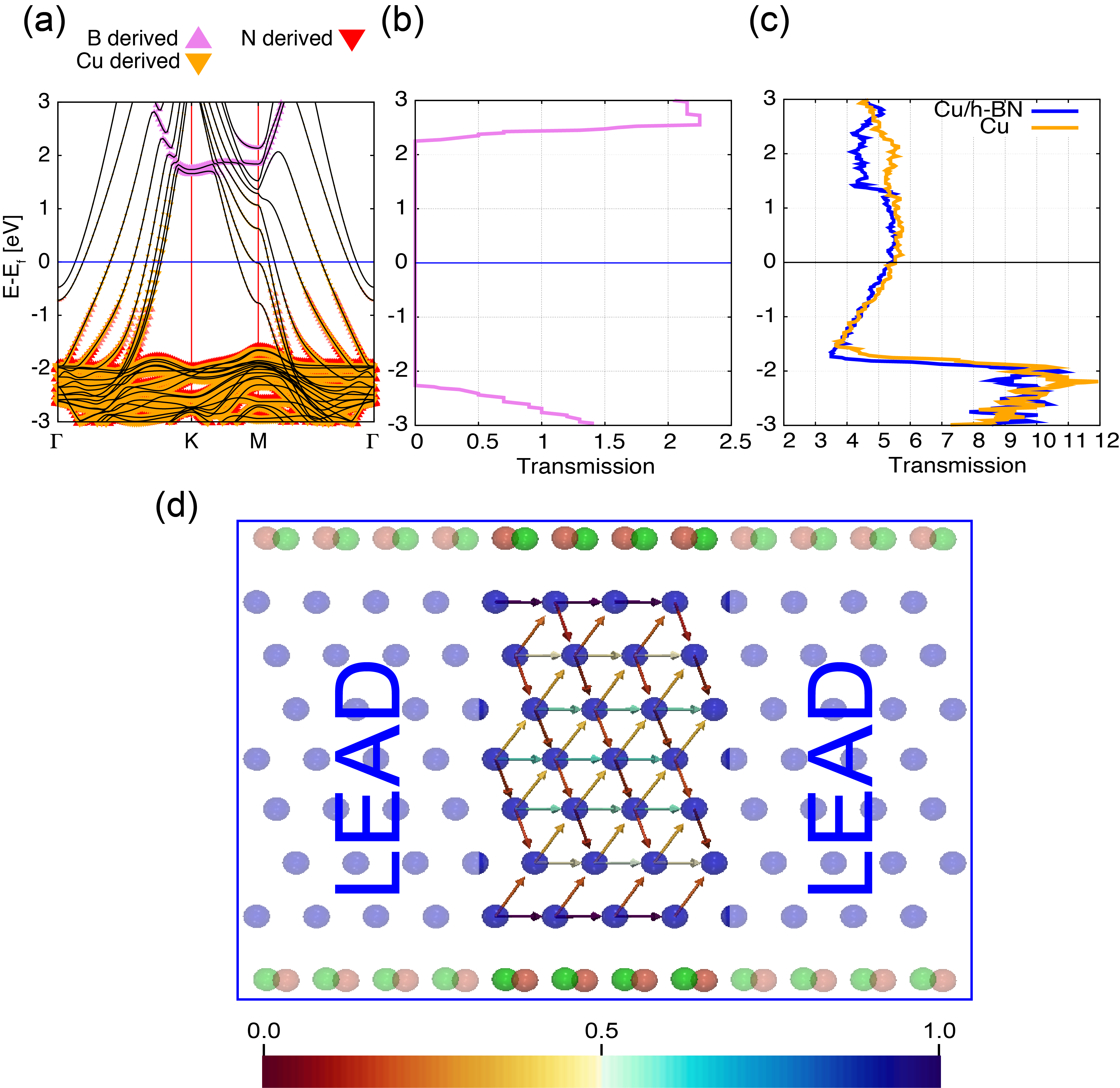}
    \caption{(a) Electronic band structure of the Cu/h-BN hybrid showing the character bands for boron (violet), nitrogen (red), and copper (orange). (b) Transmission function of h-BN. (c) Transmission function of  the Cu/h-BN hybrid (blue) and copper only (orange). (d) Local transmission at the Fermi energy of the Cu/h-BN hybrid. The color bar represents the magnitude of the transmission probability.}
\label{fig:Cu-hBN}
\end{figure}

In the case of the {\bf Copper/h-BN hybrid} the weak interactions between copper and h-BN and the fact that h-BN exhibits a large band gap, explains the lack of conducting channels at or near the Fermi level. Figure \ref{fig:Cu-hBN} shows the electronic band structure of the hybrid highlighting the boron (violet), nitrogen (red), and copper (orange) contributions to the dispersion curves. Similarly to the Cu/Gr case, the h-BN capping layers are physisorbed (Cu to graphene or h-BN distance is 
about 3.25 \AA) on the copper surfaces and as a consequence does not exhibit a strong hybridization with the copper slab as has been reported elsewhere\cite{Cuong2017,Helfrecht2017,Farmanbar2015} and electrons are expected to be confined only in the metal. \textcolor{black}{This observation is confirmed by the local transmission analysis shown in Figure \ref{fig:Cu-hBN}d where transmission pathways between the copper surface and h-BN layers are not present, similar to the case of Cu/Gr hybrids. Furthermore, due to the large band gap of h-BN and the weak interactions with the copper surfaces, there is no probable transmission paths within the h-BN layer.}

\begin{figure}[ht]
    \centering
    \includegraphics[width=0.7\textwidth]{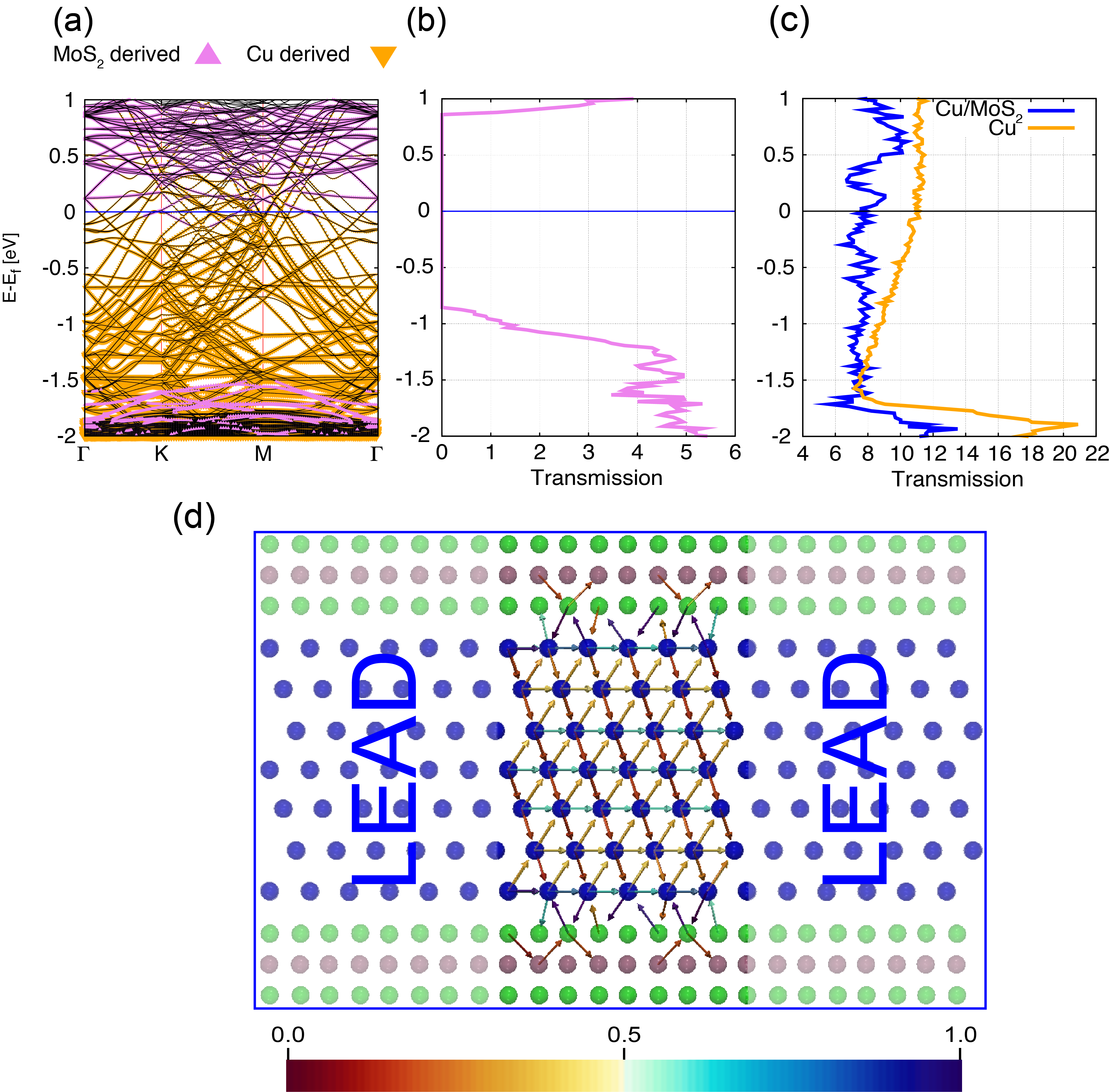}
    \caption{(a) Electronic band structure of the Cu/MoS$_2$ hybrid showing the character bands with MoS$_2$ (violet) and copper (orange) contributions. (b) Transmission function of MoS$_2$. (c) Transmission function of  the Cu/MoS$_2$ hybrid (blue) and copper only (orange). (d) Local transmission at the Fermi energy of the Cu/MoS$_2$ hybrid. The color bar represents the magnitude of the transmission probability.}
    \label{fig:Cu-MoS2}
\end{figure}

In contrast, the {\bf Copper/MoS$_2$} system shows a reduction in conductance compared to the clean copper slab. Recent \textit{ab initio} electronic structure
calculations showed that interactions go beyond van der Waals and exhibit a complex re-distribution of the electronic density within the Cu layers
near the interface and within the MoS$_2$ layers \cite{Helfrecht2017,Farmanbar2016}.
Figure \ref{fig:Cu-MoS2}a shows the electronic band dispersion of the Cu/MoS$_2$ hybrid discriminated by MoS$_2$ and copper states contributions. From the MoS$_2$ derived states shown in violet, we observe that valence and conduction bands are separated by a band gap of about 1.2 eV. \textcolor{black}{However, a detailed analysis of the character of the metallic bands crossing the Fermi level shows some contribution from the transition metal dichalcogenide layers, indicating hybridization between MoS$_2$ and copper. As a consequence, electrons can interact with the capping material (additional surface scattering) but cannot flow through it 
(due to the band gap). This is further illustrated with the local transmission analysis shown in Figure \ref{fig:Cu-MoS2}d electron transmission modes between the copper surfaces and the MoS$_2$ layers are possible with high probability. This indicates that electron density re-distribution at the Cu/MoS$_2$ interface introduces additional surface scattering. We note that the transmission pathways present within the MoS$_2$ layers do not fully span the whole central region and, most importantly, the transmission probability (magnitude) is small (about 10\%) compared to the probability of electrons hopping from the copper surface to the MoS$_2$ layers.}

\newpage

\section{Surface scattering and length Dependent Electron Transport} 
\label{sec:ElecTransII}

\begin{figure}[ht]
	\centering
	\includegraphics[width=1\textwidth]{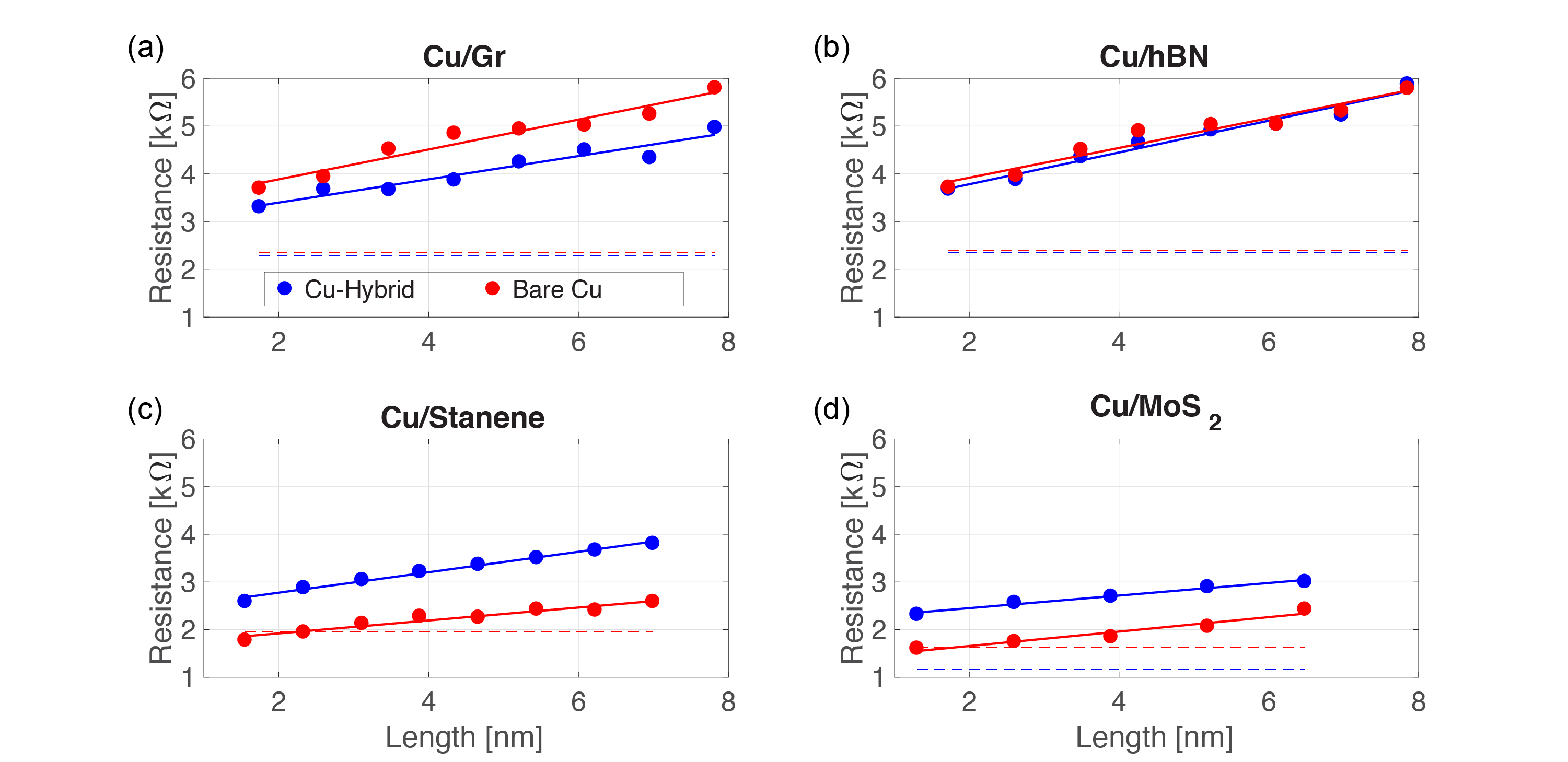}
	\caption{Resistance as functions of the slab length for (a) Cu/Gr, (b) Cu/h-BN, (c) Cu/Stanene, and (d) Cu/MoS2 with defective copper surfaces. Blue lines and resistivity correspond to that of the hybrid system and red to the defective copper slabs without capping material.}
	\label{fig:ResVsLength}
\end{figure}

The discussion so far has concentrated on ballistic transport and the effect of surface termination, we now characterize conductance
as a function of channel length to characterize how surface defects and capping affect scattering. To simulate surface disorder we follow the 
method proposed by Ke \textit{et al.}\cite{Ke2009} and characterize how resistance increases with sample length. In order to mimic surface imperfections, we introduce 25\% of Cu vacancies 
randomly distributed on both free surfaces of the copper slabs at different lengths ($L$) and then placed graphene, h-BN, stanene, and MoS$_2$ on both
free surfaces. The vacancies introduce a length dependence on the transport properties, as is shown in Figure \ref{fig:ResVsLength}. As expected, for all 
cases the resistance increases rather linearly with length given that the surface defects introduce a potential from which electrons scatter off. It is worth 
noting that our NEGF ballistic calculations capture elastic scattering with surfaces and other defects but not with phonons (all structures are minimized 
and correspond to 0K without zero point energy).  Thus, the resistance of perfect copper slab (no vacancies) is independent of length since the electron 
scattering at the surfaces is specular and the momentum in the transport direction is conserved; the calculated resistance in such a case can be interpreted 
as originating from contacts. In samples with disordered surfaces, the slope of RA with L represents a resistivity associated with elastic surface scattering. 
Our predicted value of resistivity associated with surface scattering for the copper slabs with 25\% surface vacancies range between 10-12 $\mu\Omega\cdot cm$, in good agreement with previous calculations\cite{Ke2009}.

We are interested on the effect of capping on surface scattering. Interestingly, capping with graphene on the transport properties is more significant in the case 
of defective copper surfaces than in perfect ones. We observe a decrease in resistance of approximately 20\% when capping the Cu slab containing surface 
vacancies with graphene as compared to the bare defective copper slabs, as shown in Figure \ref{fig:ResVsLength}a. 
Similar to the case of perfect copper surfaces, the effect of capping with h-BN continues to be negligible for defective surfaces due to the weak interaction 
between both materials and the lack of available conducting channels in h-BN. 
For the cases MoS$_2$ and stanene, where the interactions between the materials is stronger (see Table \ref{tab:adhesion}), the resistance of the bare 
copper thin films with surface disorder is always lower than that of the hybrid structure.      

\textbf{Surface scattering and mean free path.} The relationship between resistance and length (Figure \ref{fig:ResVsLength}) allows the characterization of the resistance per unit length (one-dimensional resistivity, $\rho_{1D}$); that is, how fast the resistance increases per unit length of conductor. Furthermore, the electron mean free path due to scattering with surface defects, $\lambda_{sd}$, can be extracted through the relation\cite{Markussen2007}:
\begin{equation*}
	R(L,E)=R_c(E)+\frac{R_c(E)}{\lambda_{sd}}L
\end{equation*}
where $R_c(E)=\frac{h}{2e^2}\frac{1}{T(E)}$ is the length independent contact resistance of the copper slab or hybrid with no surface defects and transmission $T$. The present definition of mean free path has been proven to be consistent with other formalisms\cite{Markussen2006,Avriller2006}. The one dimensional resistivity is given by $\frac{R_c(E)}{\lambda_{sd}}$, which is the slope of the linear fit to the resistance vs. length data.   

\begin{figure}[!ht]
	\centering
	\includegraphics[width=1\textwidth]{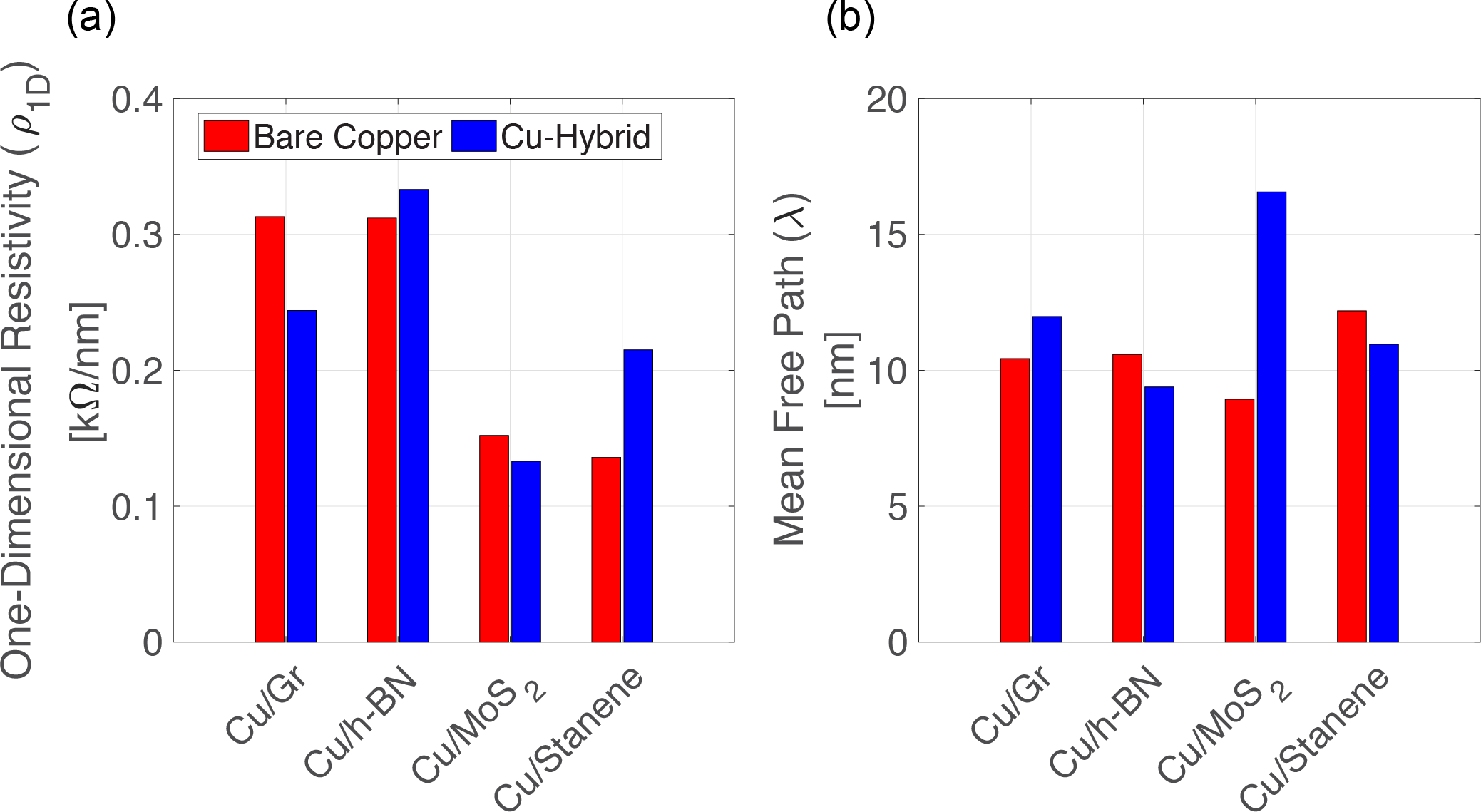}
	\caption{(a) One-dimensional resistivity of copper slabs with surface vacancies(red bars) and Cu-hybrids (blue bars). (b) Effective mean free path of electrons due to surface vacancies of copper slabs with surface vacancies (red bars) and Cu-hybrids (blue bars).}
	\label{fig:1Dres}
\end{figure}

Figure \ref{fig:1Dres}a shows the one dimensional resistivity $\rho_{1D}$ for the bare and hybrid systems with surface defects. Interestingly, we find that not 
only graphene but also MoS$_2$ reduce the $\rho_{1D}$ with respect to the bare copper thin films. This implies that capping copper thin films with graphene 
or MoS$_2$ slows down the increase of resistance as the length of the conductor increases with respect to the bare thin film. Furthermore, we  find that the 
average mean free path of electrons due to scattering with surface defects is 10 nm for bare copper thin films. Similarly to the $\rho_{1D}$ predictions, we 
observe an increase of the mean free path for Cu/Graphene and Cu/MoS$_2$ hybrids. This predictions support experimental findings where the improved 
electrical and thermal conduction of graphene encapsulated copper nanowires is attributed to an improved transmission specularity at the surfaces\cite{Mehta2015}.

\newpage

\section{Conclusion}
\label{sec:Conc}
Using DFT and NEGF, we studied the electronic transport properties of copper(111) thin films capped with two dimensional materials including graphene, hexagonal boron nitrate, molybdenum disulfide and stanene. For completeness, the transport properties of oxidized and clean copper surfaces were also characterized. For the case of Cu slabs with atomistically flat surfaces, capping with graphene increases slightly the conductance due the shift of the Fermi energy by approximately 0.75 eV above the the Dirac point of graphene; as a consequence, extra conduction channels participate in transport in Cu/Gr hybrids. The conductance of the Cu/h-BN hybrid remains almost unchanged due to the weak interactions between copper and h-BN and its wide band gap that prevents additional conducting channels near the Fermi energy of the hybrid. On the other hand, Cu/MoS$_2$ and Cu/Stanene hybrids exhibit a reduced conductance of about 30\% with respect to the clean uncapped copper slab. This is explained by the stronger interfacial interactions that results in increased surface scattering as shown by the local bond current analysis. Interestingly, we find that a thin layer of oxide on the surface of the copper films, suppresses the conductance by a factor of 10 with respect to the clean Cu surface.

Additionally, we characterized the electronic transport on defective copper surface with 25\% vacancies. We find that graphene capping decreases the resistance of the hybrid by approximately 20\% compared to the bare defective copper slab. The resistance per unit length is reduced for the defective copper surfaces capped with MoS$_2$ by 13\% with respect to the bare surface, while capping with stanene increases it by approximately 35\%. Similar to the case of clean and ideal copper surfaces, capping with h-BN does not significantly change the transport properties in defective Cu surfaces. 

\bibliography{itmdc1}

\end{document}